\documentclass[12pt]{article}
\ifx\pdfpagewidth\undefined
 \else 
\fi
\usepackage{amsmath,url,color}
\usepackage[scanall]{psfrag}
\usepackage{natbib}
\usepackage[left,displaymath,mathlines]{lineno}
\pagewiselinenumbers
 \nolinenumbers

\begin{document}
\begin{center}
{\Large\textbf{Note on the closed-form MLEs of $k$-component load-sharing systems}} \\[2ex]

\textit{Chanseok Park \\
         Department of Mathematical Sciences\\
         Clemson University \\
         Clemson, SC 29634}
\end{center}

\begin{abstract}
Consider a multiple component system connected in parallel.
In this system, as components fail one by one, 
the total load or traffic applied to the system is redistributed 
among the remaining surviving components,
which is commonly referred to as {\em load-sharing}.

Recently \cite{Kim/Kvam:2004} and \cite{Singh/Sharma/Kumar:2008} proposed 
different load-sharing models and developed parametric inference for the these models. 
However, their parametric estimates are calculated using iterative numerical methods.
In this note, we provide the general closed-form MLEs for the two load-sharing models
provided by them. 

\bigskip
\textsc{Keywords:}
Reliability, load-sharing,
maximum likelihood estimate (MLE), closed-form solution.
\end{abstract}

\clearpage
\section{Introduction}
Most research work involving load sharing models
 has mainly focused on the characterization of system reliability under a
known load-sharing rule and parameters.
The parameter estimation of the load-sharing rule has not yet been fully developed.
Recently, parametric inference for reliability under the equal load-sharing rule
has been considered by \cite{Kim/Kvam:2004} and \cite{Singh/Sharma/Kumar:2008}.
They solved the likelihood estimating equations to find the
maximum likelihood estimators (MLEs) of the load-sharing parameters. 

However, they provide no general closed form solutions for the MLEs, but instead
use iterative numerical methods to calculate their estimates. 
It is well known that there are  some problematic issues associated with  
iterative numerical methods such as stability and convergence.

In this note, we provide the general closed-form MLEs for the two load-sharing models
provided by \cite{Kim/Kvam:2004} and \cite{Singh/Sharma/Kumar:2008}. 

\section{Kim-Kvam load-sharing model}
Consider $k$-component system connected in parallel. 
Following \cite{Kim/Kvam:2004}, we assume the following:
\begin{itemize}
\item[(i)] A system is made up of $k$ components whose lifetimes are 
 independent and have identical exponential distributions with initial failure rate $\theta$. 
\item[(ii)] After the first component fails, the failure rates of the remaining 
        $k-1$ components change to $\lambda_1\theta$ where $\lambda_1>0$. 
      After the next component failure, the failure rates of the surviving $k-2$ components
      change to $\lambda_2\theta$.  Then, after the next component failure,
      the the failure rates of the surviving $k-3$ components change to 
      $\lambda_3\theta$ and so on and so forth.
\item[(iii)] There are $n$ repeated measurements of independent systems.
      That is, we have a random sample of independent systems of size $n$. 
\end{itemize}
Let $X_{im}$ denote the lifetime of the $m$-th component in the $i$-th
parallel system where $i=1,2,\ldots,n$ and $m=1,2,\ldots,k$.
For notational convenience, we can re-index the lifetimes such that
 $X_{i1} < X_{i2} < \cdots < X_{ik}$.
Then the time spacing between the $(j-1)$-th failure and
$j$-th failure for the $i$-th system is $T_{ij}=X_{ij}-X_{i,j-1}$ with
$X_{i0}=0$.

As shown in \cite{Kim/Kvam:2004},
the likelihood function for the $i$-th system is
\begin{equation*}
L_i(\theta,\Lambda)
 = (k!) \theta^k \cdot \Big[ \prod_{j=1}^{k} \lambda_{j-1} \Big] \cdot 
   \exp\Big[ -\theta \sum_{j=1}^{k} (k-j+1)\lambda_{j-1} t_{ij}  \Big], 
\end{equation*}
where $\lambda_0=1$ and $\Lambda=(\lambda_1,\ldots,\lambda_{k-1})$.  
It is immediate that the likelihood function for a random sample
of size $n$ is given by
\begin{equation} \label{EQ:Likelihood-Kim-Kvam}
L(\theta,\Lambda)
 = (k!)^n \theta^{nk} \cdot \Big[  \prod_{j=1}^{k} \lambda_{j-1}^n  \Big] \cdot 
   \exp\Big[ -\theta \sum_{i=1}^{n} \sum_{j=1}^{k} (k-j+1)\lambda_{j-1} t_{ij}  \Big].
\end{equation}
Taking the logarithm of (\ref{EQ:Likelihood-Kim-Kvam}), 
differentiating  with respect to $\theta$, 
$\lambda_1,\ldots,\lambda_{k-1}$, 
denoting partial derivative of $\log L$ with respect to $\theta$ as
$\ell_{\theta} = \partial \log L/\partial\theta$ and 
partial derivative of $\log L$ with respect to $\lambda_{j-1}$ as 
$\ell_{j-1} = \partial \log L/\partial\lambda_{j-1}$,
we obtain the log-likelihood estimating equations shown below:  
\begin{align} 
\ell_{\theta} 
 &= \frac{nk}{\theta}-\sum_{i=1}^n \sum_{j=1}^k (k-j+1)\lambda_{j-1} t_{ij}=0\label{EQ:logL0}\\
\intertext{and}
\ell_{j-1} &= \frac{n}{\lambda_{j-1}}-\theta\sum_{i=1}^n (k-j+1) t_{ij} = 0  \label{EQ:logLlam}
\end{align}
for $j=2,3,\ldots,k$. 
For convenience, we denote $t_{\bullet j} = \sum_{i=1}^{n} t_{ij}$. 
Then, (\ref{EQ:logL0}) and (\ref{EQ:logLlam}) can be rewritten as 
\begin{align}
\ell_{\theta}
&= \frac{nk}{\theta}- \sum_{j=1}^k (k-j+1)\lambda_{j-1} t_{\bullet j}=0  \label{EQ:logL01}\\
\intertext{and}
\ell_{j-1} &= \frac{n}{\lambda_{j-1}}-\theta  (k-j+1) t_{\bullet j} = 0.  \label{EQ:logLlam1}
\end{align}
It is immediate from solving (\ref{EQ:logLlam1}) for $\lambda_{j-1}$  that we have
\begin{equation} \label{EQ:lambda2}
\lambda_{j-1} = \frac{n}{\theta(k-j+1) t_{\bullet j}}, \qquad  j=2,\ldots,k. 
\end{equation}
Since $\lambda_0=1$, we rewrite (\ref{EQ:logL01}) as 
\begin{equation} \label{EQ:logL03}
\frac{nk}{\theta} - k t_{\bullet 1} - \sum_{j=2}^k (k-j+1)\lambda_{j-1} t_{\bullet j}=0.
\end{equation}
Substituting (\ref{EQ:lambda2}) into (\ref{EQ:logL03}) gives
\begin{equation} \label{EQ:logL04}
\frac{nk}{\theta} - k t_{\bullet 1} - \frac{ n(k-1) }{\theta} = 0.
\end{equation}

Solving (\ref{EQ:logL04}) for ${\theta}$, we obtain the MLE of $\theta$, denoted by 
$\hat{\theta}$,
\begin{equation} \label{EQ:MLEtheta}
\hat{\theta} = \frac{n}{k t_{\bullet 1} } = \frac{n}{\displaystyle k \sum_{i=1}^{n} t_{i1}}. 
\end{equation}
The MLEs of $\lambda_{j-1}$, denoted by $\hat{\lambda}_{j-1}$, are also easily obtained 
by substituting (\ref{EQ:MLEtheta}) into (\ref{EQ:lambda2}) 
\begin{equation*} \label{EQ:MLElambda}
\hat{\lambda}_{j-1} = \frac{k t_{\bullet 1}}{(k-j+1)t_{\bullet j}}
                    = \frac{\displaystyle k \sum_{i=1}^{n} t_{i1}}
                           {\displaystyle (k-j+1) \sum_{i=1}^{n} t_{ij}}, 
\qquad j = 2,3, \ldots, k. 
\end{equation*}

\section{Singh-Sharma-Kumar load-sharing model}
Consider $k$-component system connected in parallel. 
Following \cite{Singh/Sharma/Kumar:2008}, we assume the following:
\begin{itemize}
\item[(i)] A system is made up of $k$ components whose lifetimes are 
 independent and have exponential distributions with initial failure rate $\theta$. 
\item[(ii)] After the first component fails, 
     the failure rates of the remaining $k-1$ components 
      change to $\lambda_1\theta$ where $\lambda_1>0$. 
      After the next component failure, the failure rates of the surviving $k-2$ components
      change to $\lambda_2\theta$. 
      Then, after the next component failure, the the failure rates of the surviving $k-3$
      components change to $\lambda_3\theta$ and so on and so forth.

     After the failure of a certain number of components, say, 
     after the $s$-th component failure $(s\ge2)$, 
     the failure rates of the $k-s$ remaining components change to $\lambda_s t\theta$ 
     (linearly increasing failure rate). 
    In a similar manner, after the $(s+1)$-th component failure, the failure rates 
    of the $k-s-1$ remaining components change to $\lambda_{s+1} t\theta$, and so on. 
   Finally, after the last failure, the failure rate of the last component becomes
       $\lambda_{k-1} t\theta$. 
   
\item[(iii)] There are $n$ repeated measurements of independent systems.
      That is, we have a random sample of independent systems of size $n$. 
\end{itemize}

Again, let $X_{im}$ denote the lifetime of the $m$-th component in the $i$-th
parallel system where $i=1,2,\ldots,n$ and $m=1,2,\ldots,k$.
For notational convenience, we can re-index the lifetimes such that
 $X_{i1} < X_{i2} < \cdots < X_{ik}$.
Then the time spacing between the $(j-1)$-th failure and
$j$-th failure for the $i$-th system is $T_{ij}=X_{ij}-X_{i,j-1}$ with
$X_{i0}=0$.

As is given in \cite{Singh/Sharma/Kumar:2008}, 
the likelihood function for the $i$-th system is
\begin{align*}
L_i(\theta,\Lambda)
 =& (k!) \theta^k \cdot\Big[ \prod_{j=1}^{k} \lambda_{j-1} \Big]
                  \cdot\Big[ \prod_{j=s+1}^{k} t_{ij} \Big]   \\
  &\times 
   \exp\Big[-\theta\Big\{ \sum_{j=1}^{s}(k-j+1)\lambda_{j-1}t_{ij} 
            + \frac{1}{2} \sum_{j=s+1}^{k} (k-j+1)\lambda_{j-1}t^2_{ij} \Big\} \Big], 
\end{align*}
where $\lambda_0=1$ and $\Lambda=(\lambda_1,\ldots,\lambda_{k-1})$.  
It is immediate that the likelihood function for a random sample
of size $n$ is given by
\begin{align}  \label{EQ:Likelihood-Singh-Sharma-Kumar}
L(\theta,\Lambda)
 =& (k!)^n \theta^{nk} \cdot\Big[ \prod_{j=1}^{k} \lambda_{j-1}^n \Big]   
                       \cdot\Big[ \prod_{i=1}^{n} \prod_{j=s+1}^{k} t_{ij} \Big] \notag  \\
  &\times 
   \exp\Big[-\theta \sum_{i=1}^{n}\Big\{\sum_{j=1}^{s}(k-j+1)\lambda_{j-1}t_{ij} 
            + \frac{1}{2} \sum_{j=s+1}^{k} (k-j+1)\lambda_{j-1}t^2_{ij} \Big\} \Big].
\end{align}

Taking the logarithm of (\ref{EQ:Likelihood-Singh-Sharma-Kumar}), 
differentiating  with respect to $\theta$,
$\lambda_1,\ldots,\lambda_{k-1}$,
denoting partial derivative of $\log L$ with respect to $\theta$ as
$\ell_{\theta} = \partial \log L/\partial\theta$ and
partial derivative of $\log L$ with respect to $\lambda_{j-1}$ as
$\ell_{j-1} = \partial \log L/\partial\lambda_{j-1}$,
we obtain the log-likelihood estimating equations shown below:
\begin{align} 
\ell_{\theta} &= \frac{nk}{\theta} 
      -  \sum_{i=1}^{n}\Big\{\sum_{j=1}^{s}(k-j+1)\lambda_{j-1}t_{ij}
      + \frac{1}{2} \sum_{j=s+1}^{k} (k-j+1)\lambda_{j-1}t^2_{ij} \Big\} = 0, 
   \label{EQ:ell-theta1}  \\
\ell_{j-1} &= \frac{n}{\lambda_{j-1}} - 
          \theta \sum_{i=1}^{n} (k-j+1) t_{ij} = 0,  \qquad j=2,3,\ldots,s, 
   \label{EQ:ell-lambda-a1}  \\
\intertext{and}   
\ell_{j-1} &= \frac{n}{\lambda_{j-1}} - 
          \frac{\theta}{2} \sum_{i=1}^{n}  (k-j+1) t^2_{ij} = 0,  \qquad j=s+1,\ldots,k. 
   \label{EQ:ell-lambda-b1} 
\end{align}

Let us $y_{ij}$ define as 
\begin{equation} \label{EQ:yij}
y_{ij} = (k-j+1) t_{ij} I_{[j \le s]} + \frac{1}{2} (k-j+1) t_{ij}^2 I_{[j>s]},
\end{equation}
where $I_A$ is an indicator function whose value is one if $A$ is satisfied and
is zero if $A$ is not satisfied.  
Then (\ref{EQ:ell-theta1}), (\ref{EQ:ell-lambda-a1}) and  (\ref{EQ:ell-lambda-b1}) can be
expressed  as 
\begin{align}
\ell_{\theta} &= \frac{nk}{\theta}
      -  \sum_{i=1}^{n} \sum_{j=1}^{k} \lambda_{j-1} y_{ij} = 0, \label{EQ:ell-theta2} \\
\intertext{and}
\ell_{j-1} &= \frac{n}{\lambda_{j-1}} - \theta \sum_{i=1}^{n} y_{ij}=0,\qquad j=2,3,\ldots,k.
   \label{EQ:ell-lambda2}  
\end{align}
Notice that, by using (\ref{EQ:yij}), 
we have combined the two equations in (\ref{EQ:ell-lambda-a1}) and (\ref{EQ:ell-lambda-b1}) 
which resulted in the equation (\ref{EQ:ell-lambda2}). 

For convenience, let $y_{\bullet j} = \sum_{i=1}^{n} y_{ij}$. 
Using this and considering $\lambda_0=1$, we can rewrite 
(\ref{EQ:ell-theta2}) and (\ref{EQ:ell-lambda2}) as follows: 
\begin{align}
\ell_{\theta} &= \frac{nk}{\theta} - y_{\bullet 1} 
   - \sum_{j=2}^{k} \lambda_{j-1} y_{\bullet j} = 0, \label{EQ:ell-theta3}\\
\intertext{and}
\ell_{j-1} &= \frac{n}{\lambda_{j-1}} - \theta y_{\bullet j} = 0,  \qquad j=2,3,\ldots,k.
   \label{EQ:ell-lambda3}
\end{align}
Solving (\ref{EQ:ell-lambda3}) for $\lambda_{j-1}$, we have
\begin{equation} \label{EQ:lambda-Singh}
\lambda_{j-1} = \frac{n}{\theta y_{\bullet j}}, \qquad j=2,3,\ldots,k.
\end{equation}

Substituting (\ref{EQ:lambda-Singh}) into (\ref{EQ:ell-theta3}) gives
\begin{equation} \label{EQ:ell-theta4}
\frac{nk}{\theta} - y_{\bullet 1} - \frac{ n (k-1) }{\theta} = 0.
\end{equation}
Solving (\ref{EQ:ell-theta4}) for ${\theta}$, we obtain the MLE of $\theta$, denoted by
$\hat{\theta}$,
\begin{equation} \label{EQ:MLE-Singh-theta}
\hat{\theta} = \frac{n}{ y_{\bullet 1} } = \frac{n}{  \sum_{i=1}^{n} y_{i1}}.
\end{equation}
The MLEs of $\lambda_{j-1}$ are also easily obtained
by substituting (\ref{EQ:MLE-Singh-theta}) into (\ref{EQ:lambda-Singh})
\begin{equation*} \label{EQ:MLE-Sing-lambda}
\hat{\lambda}_{j-1} = \frac{ y_{\bullet 1}}{ y_{\bullet j}}
                    = \frac{ \sum_{i=1}^{n} y_{i1}}{ \sum_{i=1}^{n} y_{ij}},
\qquad j = 2,3, \ldots, k.
\end{equation*}

Notice that given the definition of $y_{ij}$, 
we can rewrite $y_{\bullet j}$ in the following manner:
$$
y_{\bullet j} =
\left\{ \begin{array}{r@{\qquad}l}
   \displaystyle\sum_{i=1}^n (k-j+1) t_{ij} = (k-j+1) \sum_{i=1}^n t_{ij}, & j=2,3,\ldots,s \\
   \displaystyle\frac{1}{2} \sum_{i=1}^n (k-j+1) t^2_{ij} 
       = \frac{1}{2} (k-j+1) \sum_{i=1}^n t^2_{ij},  & j=s+1,\ldots,k   \\
\end{array} \right. .
$$
Therefore we can write the closed form MLEs for $\theta$ and $\lambda_{j-1}$ as
\begin{align*}
\hat{\theta} &= \frac{n}{\displaystyle k \sum_{i=1}^{n} t_{i1}}  \\
\intertext{and} 
\hat{\lambda}_{j-1} &=  
\left\{ \begin{array}{r@{\qquad}l}
   \frac{\displaystyle k \sum_{i=1}^{n} t_{i1}}
        {\displaystyle (k-j+1) \sum_{i=1}^n t_{ij}}, & j=2,3,\ldots,s   \\
   \frac{\displaystyle k \sum_{i=1}^{n} t_{i1}}
        {\displaystyle \frac{1}{2} (k-j+1) \sum_{i=1}^n t^2_{ij} }, & j=s+1,\ldots,k   \\
\end{array} \right..
\end{align*}




\end{document}